\documentclass[aps,pra,reprint,superscriptaddress,showkeys]{revtex4-2}
\usepackage[T1]{fontenc}
\usepackage{amsmath,amssymb,amsfonts,amsthm}
\usepackage{mathtools}
\usepackage{siunitx}
\usepackage{graphicx}
\usepackage[colorlinks=true,allcolors=blue]{hyperref}
\usepackage[caption=false,labelformat=empty,subrefformat=parens]{subfig}
\usepackage{xcolor}
\graphicspath{{figures/}}

\definecolor{DB}{rgb}{0.8,0.00,0.00}

\begin{document}
\date{\today}
\author{Daniel Bruns} \email{dbruns@phas.ubc.ca}
\affiliation{Department of Physics and Astronomy, University of
  British Columbia, Vancouver, BC, Canada V6T~1Z1}
\affiliation{Quantum Matter Institute, University of British Columbia,
  Vancouver, BC, Canada V6T~1Z4} \author{Alireza Nojeh}
\affiliation{Department of Electrical and Computer Engineering,
  University of British Columbia, Vancouver, BC, Canada V6T~1Z4}
\affiliation{Quantum Matter Institute, University of British Columbia,
  Vancouver, BC, Canada V6T~1Z4} \author{A.~Srikantha Phani}
\affiliation{Department of Mechanical Engineering, University of
  British Columbia, Vancouver, BC, Canada V6T~1Z4} \author{J\"org
  Rottler} \affiliation{Department of Physics and Astronomy,
  University of British Columbia, Vancouver, BC, Canada V6T~1Z1}
\affiliation{Quantum Matter Institute, University of British Columbia,
  Vancouver, BC, Canada V6T~1Z4}

\title{Nanotube heat conductors under tensile strain:\\ Reducing the
  three-phonon scattering strength of acoustic phonons}

\begin{abstract}
  Acoustic phonons play a special role in lattice heat transport, and
  confining these low-energy modes in low-dimensional materials may
  enable nontrivial transport phenomena. By applying lowest-order
  anharmonic perturbation theory to an atomistic model of a carbon
  nanotube, we investigate numerically and analytically the spectrum
  of three-phonon scattering channels in which at least one phonon is
  of low energy. Our calculations show that acoustic longitudinal
  (LA), flexural (FA), and twisting (TW) modes in nanotubes exhibit a
  distinct dissipative behavior in the long-wavelength limit,
  $|k| \rightarrow 0$, which manifests itself in scattering rates that
  scale as $\Gamma_{\rm{LA}}\sim |k|^{-1/2}$,
  $\Gamma_{\rm{FA}}\sim k^0$, and $\Gamma_{\rm{TW}}\sim
  |k|^{1/2}$. These scaling relations are a consequence of the
  harmonic lattice approximation and critically depend on the
  condition that tubes are free of mechanical strain. In this regard,
  we show that small amounts of tensile lattice strain $\epsilon$
  reduce the strength of anharmonic scattering, resulting in
  strain-modulated rates that, in the long-wavelength limit, obey
  $\Gamma \sim \epsilon^{r} |k|^{s}$ with $r\leq 0$ and $s\geq 1$,
  irrespectively of acoustic mode polarization.  Under the single-mode
  relaxation time approximation of the linearized Peierls-Boltzmann
  equation (PBE), the long-tube limit of lattice thermal conductivity
  in stress-free and stretched tube configurations can be
  unambiguously characterized. Going beyond relaxation time
  approximations, analytical results obtained in the present study may
  help to benchmark numerical routines which aim at deriving the
  thermal conductivity of nanotubes from an exact solution of the PBE.

\end{abstract}

\maketitle


\section{Introduction}
\label{sec:introduction}

Two-dimensional graphene sheets and their one-dimensional derivatives,
carbon nanotubes, exhibit an extremely efficient lattice heat
transport~\cite{Balandin2008,Pop2006}. While the merits of integrating
these carbon allotropes into thermal management applications have been
convincingly demonstrated in recent
years~\cite{Marconnet2011a,Shahil2012,Yan2012,Kholmanov2015,Liu2016,Dai2019},
separating their intrinsic low-dimensional heat transport properties
from external factors remains a
challenge~\cite{B11,Balandin2020}. Residual strain in the carbon
lattice is commonly identified as one source of measurement
uncertainty among others in heat transport experiments involving
nanostructured carbon samples~\cite{B11,Balandin2020}. A thorough
understanding on theoretical grounds of the role of strain in
low-dimensional lattice heat transport is therefore crucial.

Focusing on the effects of strain in lattice heat transport, phonon
frequencies in three-dimensional crystals are typically described by a
positive mode-specific Gr\"uneisen parameter~\cite{Xu1991,Mounet2005},
which translates to frequency hardening of phonons under compressive
and softening under tensile loading. Roughly speaking, an overall
downward shift of frequencies under a tensile load leads to a
reduction of phonon group velocities, which in turn causes a reduced
rate of lattice heat transfer. Even though other phonon properties
play a role as well, such a qualitative trend was generally confirmed
for three-dimensional systems~\cite{Parrish2014}. In contrast, one and
two-dimensional crystals allow for atomic out-of-line and out-of-plane
vibrations, respectively, whose associated phonon frequencies instead
harden under tensile strain~\cite{Mounet2005,ADE2013}. In light of
these \emph{flexural} modes, nonconventional strain behavior of
low-dimensional heat transport in graphene and carbon nanotubes might
be anticipated.

The linearized Peierls-Boltzmann equation (PBE) for phonon transport
has become a standard tool to tackle anharmonic phonon interactions
and the emergence of thermal resistivity in crystalline
solids~\cite{McGaughey2019}. Within this formalism, determining the
kinematically allowed three-phonon processes and their corresponding
scattering amplitudes, as prescribed by anharmonic perturbation
theory, is a key requirement for predicting the thermal
conductivity. Bonini et al.~\cite{BGM2012} studied graphene in the
framework of the PBE, making the notable observation that anharmonic
scattering processes including acoustic flexural modes become
systematically weakened under increasing tensile lattice
strain. Specifically, by examining analytically the implications of
anharmonic perturbation theory in the limit of low phonon energies, it
was shown in Ref.~\cite{BGM2012} that the per-phonon conductivity
contributions of weakly damped flexural modes under strain might grow
without bound in the limit of large crystalline domain sizes. Although
these conclusions with respect to thermal transport in graphene were
later called into question and attributed to an inadequacy of
relaxation time approximations to the
PBE~\cite{Fugallo2014,Cepellotti2016}, the results in
Ref.~\cite{BGM2012}, nevertheless, give an unequivocal demonstration
of the importance of lattice strain as far as low-frequency phonon
scattering rates are concerned.

In the case of carbon nanotubes, some analytical considerations of
three-phonon scattering rates were first presented by Mingo and
Broido~\cite{MB05}. By simplifying the exact scattering formulas
inherent to anharmonic perturbation theory, they derived approximate
long-wavelength scaling relations capturing the three-phonon
scattering strength of acoustic phonons, and thereby pointed out the
issue of heat carrying modes with vanishing scattering rates. Later,
more refined heat transport calculations within the PBE framework were
performed on free-standing carbon nanotubes by us~\cite{Bruns2020} and
others~\cite{LBM09,LBM10,YOH15,Barbalinardo21}. With the help of
numerical routines, these studies retained the full complexity
associated with the many-body problem of interacting phonons, but did
not account for the \emph{exact} long-wavelength scaling relations of
acoustic phonon scattering rates which are prescribed by lowest-order
anharmonic perturbation theory. More so, to the best of our knowledge,
the effects of strain on low-frequency three-phonon scattering rates
involving flexural out-of-line vibrations have so far not been
rigorously addressed within the PBE formalism.

To shed light on the manifestations of strain in lattice heat
transport under one-dimensional phonon confinement, we examine here
the case of carbon nanotubes in stress-free and stretched
configurations. In particular, in anticipation of anomalous strain
behavior rooted in the dynamics of flexural acoustic
phonons~\cite{BGM2012}, we put our emphasis on examining the
three-phonon scattering amplitudes of nanotubes in the low-frequency
limit. To this end, we carry out lattice-dynamical calculations which
allow us to produce some general long-wavelength scaling relations
resulting from lowest-order anharmonic perturbation theory, which are
demonstrated to depend crucially on both tensile lattice strain as
well as acoustic mode polarization. Importantly, while previous
studies on stress-free
nanotubes~\cite{LBM09,LBM10,YOH15,Bruns2020,Barbalinardo21}
resorted exclusively to reporting numerical results, here we support
our data with an exact analysis of the behavior of three-phonon
scattering rates in the low-frequency limit. As we show, earlier
continuum modeling efforts on
nanotubes~\cite{Popov2000,SA2002,MEG2009,GK2019} can, in part, help
put numerical lattice-dynamical findings on firm theoretical footing.
In this way, at the level of the single-mode relaxation time
approximation to the PBE, unambiguous qualitative trends of the
thermal conductivity in the long-tube limit can be given for both
unstrained and stretched tube configurations.

This paper is structured as follows. In
Sec.~\ref{sec:latt-dynam-calc}, we give a brief account of our
lattice-dynamical model and the numerical treatment of three-phonon
interactions, with further computational details deferred to
Appendix~\Ref{sec:three-phon-coupl}. Section~\ref{sec:continuum-models}
establishes relevant links to continuum theories. Strain dependent
scattering rates of acoustic phonons are presented in
Sec.~\ref{sec:mode-level-analysis}, and are complemented with exact
long-wavelength scaling laws derived in
Appendix~\ref{sec:asympt-scal-laws}. In Sec.~\ref{sec:therm-cond-},
these findings are applied to predict thermal conductivity
coefficients of pristine tubes under a relaxation time approximation
to the PBE. We finally discuss further research directions in
Sec.~\ref{sec:discussion} and conclude in Sec.~\ref{sec:conclusion}.

\section{Lattice-dynamical calculations}
\label{sec:latt-dynam-calc}

To model carbon nanotubes, throughout this work we adopt a Tersoff
type atomic interaction potential devised for sp2-hybridized
carbon~\cite{LB10}. Rather than aiming at precise quantitative
predictions accounting for tube chirality effects, our focus lies on
finding generic trends of acoustic mode interactions in light of
one-dimensional phonon confinement and tensile lattice strain. We
therefore restrict our attention to isotropically pure nanotubes of
chirality (4,4) with diameter $D\approx \SI{5.55}{\angstrom}$ and
translational lattice parameter $a=(1+\epsilon)a_0$, where
$\epsilon\geq 0$ denotes a variable tensile strain amplitude and
$a_0\approx \SI{2.51}{\angstrom}$ corresponds to the tube's
minimum-potential energy configuration in the stress-free state. Such
achiral tubes of small diameter give rise to only a moderate number of
phonon dispersion branches $\omega_{j}(k)$, in this case
$j=1,\dots,48$. As a result, it is possible to consider the full
spectrum of three-phonon scattering processes over a relatively fine
wave number grid, spanning the one-dimensional Brillouin zone
$-\pi/a< k \leq \pi/a$, which, as we show later, is crucial to resolve
acoustic phonon dynamics in the long-wavelength limit $|k|a\ll 1$.

We calculate anharmonic phonon-phonon scattering rates within
lowest-order perturbation theory. For a one-dimensional system, the
total rate for a mode with polarization $j$ and wave number $k$ due to
three-phonon interactions is derived, e.g., from the imaginary part of
the phonon self-energy~\cite{MF62} and expressed as
\begin{align}
  \label{eq:1}
  \begin{split}
    \Gamma_{j}(k)=\frac{1}{N}\sum_{\substack{(k',j')\\(k'',j'')}}
    \frac{1}{2}&\Gamma^{-}_{jj'j''}(k,k',k'')
    \\
    &+\, \Gamma^{+}_{jj'j''}(k,k',k''),
  \end{split}
\end{align}
where $N$ determines the grid spacing of wave numbers,
$\Delta k=2\pi/Na$, and the $\Gamma^{\pm}$'s describe three-phonon
absorption ($+$) as well as decay ($-$) amplitudes. Introducing the
shorthand $\nu=(j,k)$, individual transition amplitudes at finite
temperature $(T>0)$ are given by
\begin{align}
  \label{eq:2}
  \begin{split}
    \Gamma^{\pm}_{\nu,\nu',\nu''}&=\frac{\hbar
      \pi}{4\,\omega_{\nu}\omega_{\nu'}\omega_{\nu''}}\frac{\left(n_{\nu'}+1/2\pm
        1/2\right)n_{\nu''}}{n_{\nu}}|V^{\pm}_{\nu,\nu',\nu''}|^2\\
    &\times \delta(\omega_{\nu}\pm
    \omega_{\nu'}-\omega_{\nu''})\Delta_{k\pm k'-k'',\,2\pi m/a},
  \end{split}
\end{align}
where the $n_{\nu}$'s are the Bose-Einstein occupation numbers,
$V^{\pm}_{\nu,\nu',\nu''}=V_{jj'j''}(k,\pm k',-k'')$ denote the
Fourier transforms of the third order force constant tensor (see
Appendix~\ref{sec:three-phon-coupl}), the delta function enforces
energy conservation, and the discrete translational invariance of the
crystal lattice implies the conservation of quasimomentum as
signified by the Kronecker delta $\Delta$, with $m=0$ and $m=\pm 1$
denoting normal and Umklapp scattering processes, respectively. Given
the discrete rotational invariance of nanotubes, scattering processes
are further subject to a quasiangular momentum selection
rule~\cite{LBM09,Lindsay2012}, which is not explicitly stated here but
encoded in $V^{\pm}_{\nu,\nu',\nu''}$. Phonon triplets sampled from
the same regular wave number grid generally cannot fulfill the
condition of energy conservation and the delta function in
Eq.~\eqref{eq:2} has to be resolved numerically. Here, we apply an
adaptive Gaussian smearing approach~\cite{LCK14} to compute
phonon-phonon scattering rates as per
Eqs.~\eqref{eq:1} and~\eqref{eq:2}. Tensile lattice strain $\epsilon$
enters naturally into our calculations by influencing second and third
order atomic force constants, which are the essential inputs to
predict harmonic phonon spectra and three-phonon coupling coefficients
$V_{jj'j''}$.

\section{Continuum models\\ and acoustic phonons in nanotubes}
\label{sec:continuum-models}

\begin{figure}
  \includegraphics[width=0.95\columnwidth]{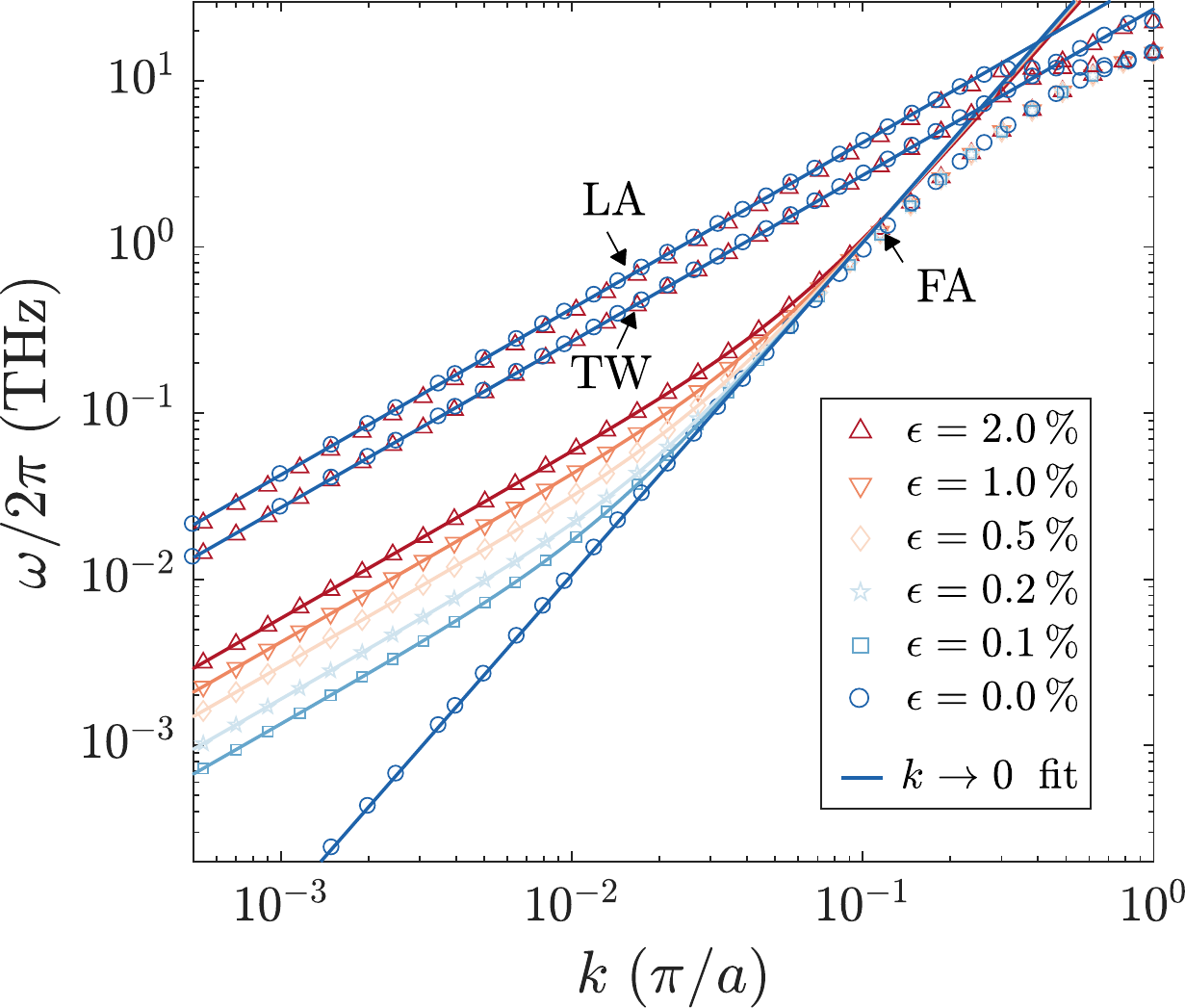}
  \caption{Log-log plot of the acoustic phonon dispersion in (4,4)
    carbon nanotubes under varying tensile strain $\epsilon$. Empty
    plot markers correspond to lattice-dynamical calculations that
    take strain dependent second order force constants as input. Solid
    lines are obtained by fitting the continuum predictions as per
    Eqs.~\eqref{eq:3} and~\eqref{eq:4} to phonon data in the region
    $k\leq 0.1\,\pi/a$. Here, $\omega_{\rm{TW}}=27.04\,k$,
    $\omega_{\rm{LA}}=42.56\,k$ and
    $\omega^2_{\rm{FA}}=(10.61-73.70\epsilon)\times 10^3\, k^4 +1.75
    \times 10^3\,\epsilon\,k^2$ for twisting (TW), longitudinal (LA),
    and flexural (FA) modes, respectively, with $\omega_j$ in units
    of $\rm{THz}$ and $k$ in units of $\pi/a$, where $a$ denotes the
    lattice constant at a given strain amplitude.}\label{fig:1}
\end{figure}

Treated as isotropic, free-standing, linear elastic continua, both
long tubes and solid rods of radius $R$ are known to permit three
distinctive types of low-frequency waves: twisting (TW), longitudinal
(LA) and two degenerate flexural bending (FA) type modes, exhibiting
linear, $\omega_{\rm{TW/LA}}\sim |k|$, and quadratic,
$\omega_{\rm{FA}}\sim R k^2$, axial wave dispersion in the
long-wavelength limit $|k|R\ll 1$~\cite{Viola2007}.  In the case of
carbon nanotubes, these acoustic waves can be reproduced by a simple
continuum model of a cylindrical
surface~\cite{SA2002,Popov2000}. Specifically, by assuming that carbon
nanotubes inherit the in-plane elastic isotropy of graphene
sheets~\cite{SA2002,Popov2000}, the acoustic wave dispersion for
$|k|\rightarrow 0$ becomes
\begin{align}
  \label{eq:3}
  \begin{split}
    \omega_{\rm{TW}}(k)&=v_{\rm{TW}}|k|,\quad v_{\rm{TW}}=\sqrt{\frac{\mu}{\rho_{m}}},\\
    \omega_{\rm{LA}}(k)&=v_{\rm{LA}}|k|,\quad v_{\rm{LA}}=\sqrt{\frac{Y}{\rho_m}},\\
    \omega_{\rm{FA}}(k)&=\frac{v_{\rm{LA}}R}{\sqrt{2}}k^2,
  \end{split}
\end{align}
where the TW and LA wave velocities are determined by the 2D shear
modulus $\mu$, the 2D Young's modulus $Y$, and the area mass density
of the cylinder surface $\rho_{m}$. More sophisticated continuum
models have to be employed to account for tube chirality effects and
elastic anisotropy.  Considering both chiral and achiral tubes,
recently, Gupta and Kumar~\cite{GK2019} applied a Cosserat rod theory
in order to derive analytical expressions for the acoustic wave
dispersion under external loads. Following Ref.~\cite{GK2019}, to the
lowest order in the tensile stain amplitude $\epsilon$, frequency
hardening of FA modes manifests itself in nanotubes as
\begin{align}
  \label{eq:4}
  \omega^2_{\rm{FA}}(k,\epsilon)=\omega^2_{\rm{FA}}(k,0)+\mathcal{T}_1\epsilon\,k^2-\mathcal{T}_2\epsilon\,k^4,
\end{align}
where $\omega_{\rm{FA}}(k,0)$ denotes the FA mode dispersion in the
stress-free state and the $\mathcal{T}_{i}$'s stand for some positive
constants which relate to the tube's stretching and bending
stiffness. For our model of a (4,4) carbon nanotube, acoustic mode
frequencies obtained from lattice-dynamical calculations are plotted
in Fig.~\ref{fig:1}. From TW and LA mode frequencies in unstrained
tubes, we infer $v_{\rm{TW}}=\SI{13.57}{km/s}$ and
$v_{\rm{LA}}=\SI{21.37}{km/s}$, respectively. At
$\epsilon=\SI{2.0}{\%}$, TW mode velocities are found to be slightly
increased by $\SI{0.2}{\%}$, whereas LA mode velocities become smaller
by $\SI{2.9}{\%}$. Most importantly, under finite tensile strain,
long-wavelength FA modes transition from a quadratic to a linear wave
dispersion as is faithfully described by Eq.~\eqref{eq:4}.

Continuum models might be further invoked to provide insight into
acoustic phonon-phonon interaction processes. Within lowest-order
perturbation theory, for a given acoustic phonon $(j,k)$ of low energy
and long wavelength, $\omega_j(|k|\rightarrow 0)=0$, two types of
interacting phonon triplets are permitted by the law of energy
conservation: (i) triplets of three acoustic phonons, whose energies
fall below the lowest optical phonon energy,
$\omega_{j}\sim\omega_{j'}\sim \omega_{j''}\ll \omega_{\rm{OPT}}$, or
(ii) triplets of one acoustic and two optical modes, where
$\omega_j\ll \omega_{\rm{OPT}'}\sim \omega_{\rm{OPT}''}$. A continuum
approach lends itself to study scattering channels involving triplets
of type (i). Treating stress-free carbon nanotubes as isotropic hollow
cylinders and applying tools of nonlinear elasticity
theory~\cite{L.D1986}, it was shown by De Martino et
al.~\cite{MEG2009} that possible interacting acoustic mode triplets
reduce to $(\rm{LA},\rm{LA},\rm{LA})$, $(\rm{LA},\rm{TW},\rm{TW})$,
$(\rm{LA},\rm{FA},\rm{FA})$, and $(\rm{TW},\rm{FA},\rm{FA})$ for which
coupling coefficients $V_{jj'j''}(k,k',k'')$ can be derived in terms
of second and third order elastic constants. In particular, a
noteworthy result of Ref.~\cite{MEG2009} is the quartic wave number
dependence $V_{\rm{TWFAFA}}\sim kk'k''(k'-k'')R$, which is at variance
with the standard cubic long-wavelength approximation
$V_{jj'j''}\sim kk'k''$ for type (i) triplets commonly found in
classic textbooks~\cite{Z60,Pitaevskii1981}. Within a
lattice-dynamical treatment of three-phonon interactions, this
behavior in carbon nanotubes was previously
overlooked~\cite{LBM09}. As we show below, the weak coupling strength
between TW and FA modes gives rise to an intricate scattering behavior
of low-frequency TW modes in stress-free and stretched tubes.

\section{Acoustic mode level analysis}
\label{sec:mode-level-analysis}

Taking into account the full spectrum of three-phonon interactions
including type (i) and (ii) phonon triplets, hereinafter, we evaluate
acoustic phonon scattering rates within our lattice-dynamical model
according to Eqs.~\eqref{eq:1} and~\eqref{eq:2} as a function of
tensile strain $\epsilon$. Specifically, in order to infer the
governing long-wavelength scaling relations
$\Gamma_{j}\sim \epsilon^{r}k^{s}$ for
$j\in\{\rm{LA},\rm{FA},\rm{TW}\}$, we decompose wave number-resolved
scattering rates in the long-wavelength limit $0<ka\ll 1$ into
contributions that stem from individual three-phonon decay and
absorption channels,
$\Gamma_{j}=\sum_{j',j''}\Gamma^{-}_{jj'j''}/2+\Gamma^{+}_{jj'j''}$.
For brevity in notation, mode polarizations appearing in triple
subscripts are henceforth signified by single letters. For example, a
triplet of type (ii) involving LA modes will be represented by
$\rm{LOO}$.

\subsection{Longitudinal modes}
\label{sec:longitudinal-modes}

\begin{figure*}
  \includegraphics[width=\textwidth]{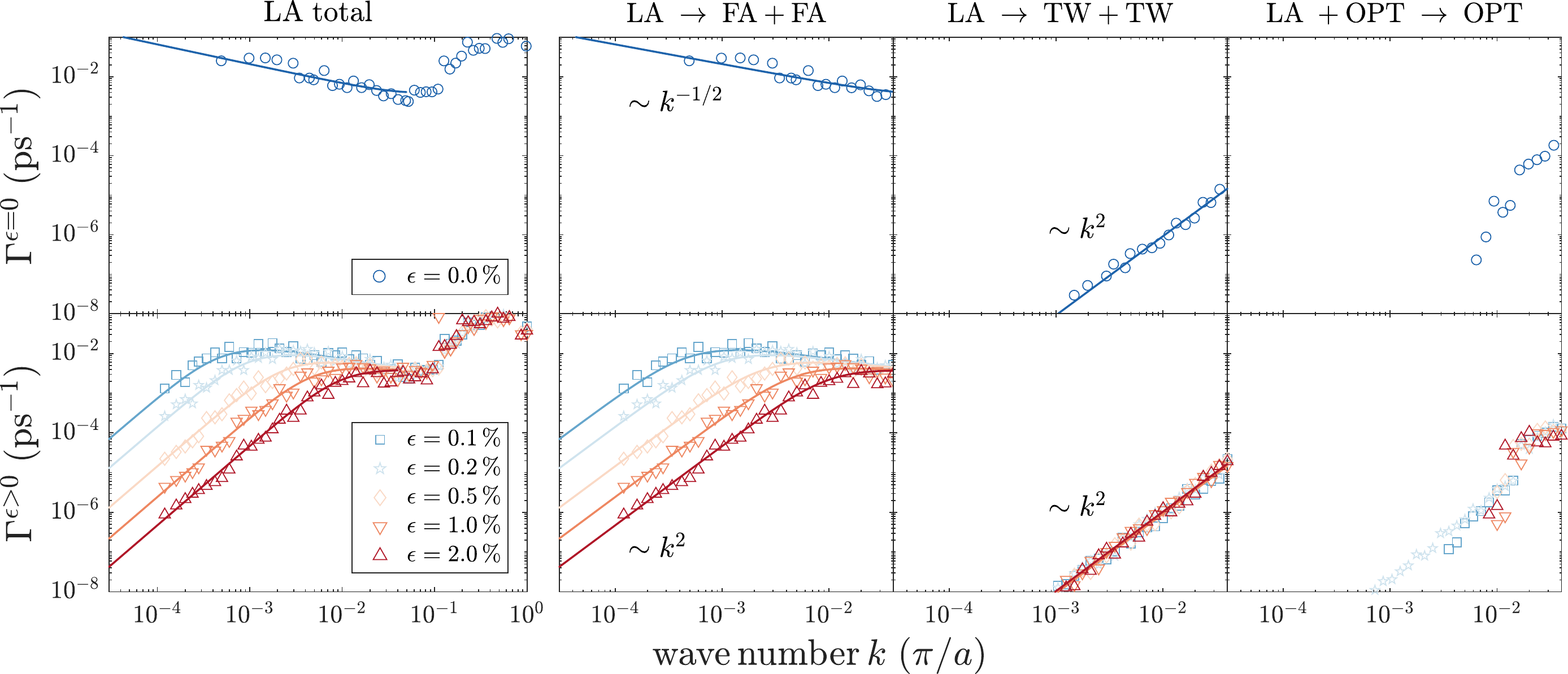}
  \caption{Anharmonic scattering rates (left) of LA modes at room
    temperature and decomposition (right) into individual three-phonon
    scattering channels in the region $k\leq 0.04\, \pi/a$. Top and
    bottom panels show stress-free, $\epsilon=0$, and stretched,
    $\epsilon>0$, configurations of a (4,4) carbon nanotube,
    respectively. Lattice-dynamical predictions as per
    Eqs.~\eqref{eq:1} and~\eqref{eq:2} are represented by empty plot
    markers. Solid curves are obtained by substituting long-wavelength
    approximations into Eqs.~\eqref{eq:1} and~\eqref{eq:2}. In the
    $ka\ll 1$ limit, the total LA scattering rate is dominated by the
    decay process $\rm{LA} \rightarrow \rm{FA} + \rm{FA}$. Induced by
    strain, frequency hardening of FA modes causes total scattering
    rates to transition from a scaling $\sim k^{-1/2}$ in stress-free
    tubes to a scaling $\sim \epsilon^{-5/2} k^{2}$ in strained
    tubes.}
  \label{fig:2}
\end{figure*}

Given the large group velocity of LA modes, selection rules imposed by
energy and quasimomentum conservation in Eq.~\eqref{eq:2} imply a
severely reduced three-phonon scattering phase space available to
long-wavelength LA modes. In Fig.~\ref{fig:2}, we show total LA
scattering rates under varying tensile strain amplitudes at different
wavelength scales. Decomposing our data for small $k$, we detect three
different types of scattering channels that contribute to the total LA
rate in the long-wavelength limit:
$\rm{LA}\rightarrow \rm{FA}+\rm{FA}$,
$\rm{LA}\rightarrow \rm{TW}+\rm{TW}$, and
$\rm{LA}+\rm{OPT}\rightarrow \rm{OPT}$ among which the first dominates
in both unstrained and strained tubes.

In the absence of lattice strain, one finds
$\Gamma^{\epsilon=0}_{\rm{LA}}\sim\Gamma^{-}_{\rm{LFF}}\sim
k^{-1/2}$ which can be traced back to the long-wavelength scaling
relations $\omega_{\rm{FA}}\sim k^2$ and $V_{\rm{LFF}}\sim kk'k''$
(see Appendix~\ref{sec:decay-into-two}).
As can be noted from the top left-hand panel of Fig.~\ref{fig:2}, the
square root scaling breaks down at smaller wavelength scales
$ka > 0.1\pi$. This threshold conforms with the phonon dispersion data
in Fig.~\ref{fig:1}, where it can been seen that FA mode frequencies
deviate from the quadratic trend once $ka > 0.1 \pi$, indicating a
regime where atomistic details entering into the exact phonon
dispersion become important.

In the presence of tensile lattice strain, $\epsilon>0$,
lattice-dynamical predictions in Fig.~\ref{fig:2} indicate a reduction
of long-wavelength rates with increasing strain amplitude and a
transition away from the square root scaling in unstrained tubes. In
order to explain the strain dependence of $\Gamma_{\rm{LA}}$, we
resolve $\Gamma^{-}_{\rm{LFF}}$ as per
Eqs.~\eqref{eq:1} and~\eqref{eq:2} in the continuum limit
$\sum_{k}\rightarrow \int {\rm{d}}k$ which requires knowledge of the
functional wave number dependencies of $\omega_{\rm{LA}}$,
$\omega_{\rm{FA}}$ and $V_{\rm{LFF}}$. Here, we adopt the
dispersion fit lines of Fig.~\ref{fig:1} and, under the assumption
that strain has no effect on the cubic wave number dependence of
three-phonon couplings, we set
$V_{\rm{LFF}}=C_{3}\,kk'k''$. Representing the three-phonon
coupling strength, the only unknown parameter $C_{3}$ is then fitted
to lattice-dynamical data of $\Gamma^{-}_{\rm{LFF}}$. This
procedure yields the solid lines in Fig.~\ref{fig:2} for tubes in
unstrained and strained configurations. In fitting our
lattice-dynamical predictions, we do not observe a systematic
variation of $C_{3}$ under varying strain amplitudes,
$C_{3}\nsim \epsilon$, which leaves $\omega_{\rm{FA}}$ as the only
strain dependent quantity entering into
$\Gamma^{-}_{\rm{LFF}}$. Thus, the strain induced downward shift of
LA rates can be solely attributed to the frequency hardening of FA
modes.

The implications of a linearized FA mode dispersion are best
understood in the phonon frequency domain, where the density of FA
states $D_{\rm{FA}}$ within the low-frequency interval
$\omega+{\rm{d}}\omega$ goes from $D_{\rm{FA}}\sim \omega^{-1/2}$ in
unstrained to $D_{\rm{FA}}\sim \omega^{0}$ in strained
systems. Loosely speaking, under increasing tensile lattice strain,
there are less and less preferential FA states that a given
long-wavelength LA mode can decay into. Assuming a strictly linear FA
mode dispersion with strain dependent prefactor,
$\omega_{\rm{FA}}\sim \epsilon^{1/2}|k|$, one finds
$\Gamma^{\epsilon>0}_{\rm{LA}}\sim \Gamma^{-}_{\rm{LFF}}\sim
\epsilon^{-5/2}k^2$ (see Appendix \ref{sec:decay-into-two}).

The scattering amplitudes $\Gamma^{-}_{\rm{LTT}}$ and
$\Gamma^{+}_{\rm{LOO}}$, both of which do not involve
long-wavelength FA modes, remain relatively insensitive to strain. A
scaling behavior $\Gamma^{-}_{\rm{LTT}}\sim k^2$ is deducible from
a linear dispersion $\omega_{\rm{TW}}\sim |k|$ and a coupling
coefficient $V_{\rm{LTT}}\sim kk'k''$ (see Appendix
\ref{sec:onel-two-twist}). To produce the solid lines in
Fig.~\ref{fig:2} for $\Gamma^{-}_{\rm{LTT}}$, as before, the
unknown prefactor of $V_{\rm{LTT}}$ is taken as a fit parameter to
our lattice-dynamical predictions. In the case of optical absorption
processes, numerical noise inherent to our evaluation of
Eqs.~\eqref{eq:1} and~\eqref{eq:2} becomes apparent to the extent that no
continuous trend lines for $\Gamma^{+}_{\rm{LOO}}$ are
discernable. Notwithstanding the uncertainty in our prediction of
$\Gamma^{+}_{\rm{LOO}}$, data points in Fig.~\ref{fig:2} suggest
that optical absorption processes as well as TW mode interactions
contribute negligibly to the dissipation of long-wavelength LA modes.

\subsection{Flexural modes}
\label{sec:flexural-modes}

\begin{figure*}
  \includegraphics[width=\textwidth]{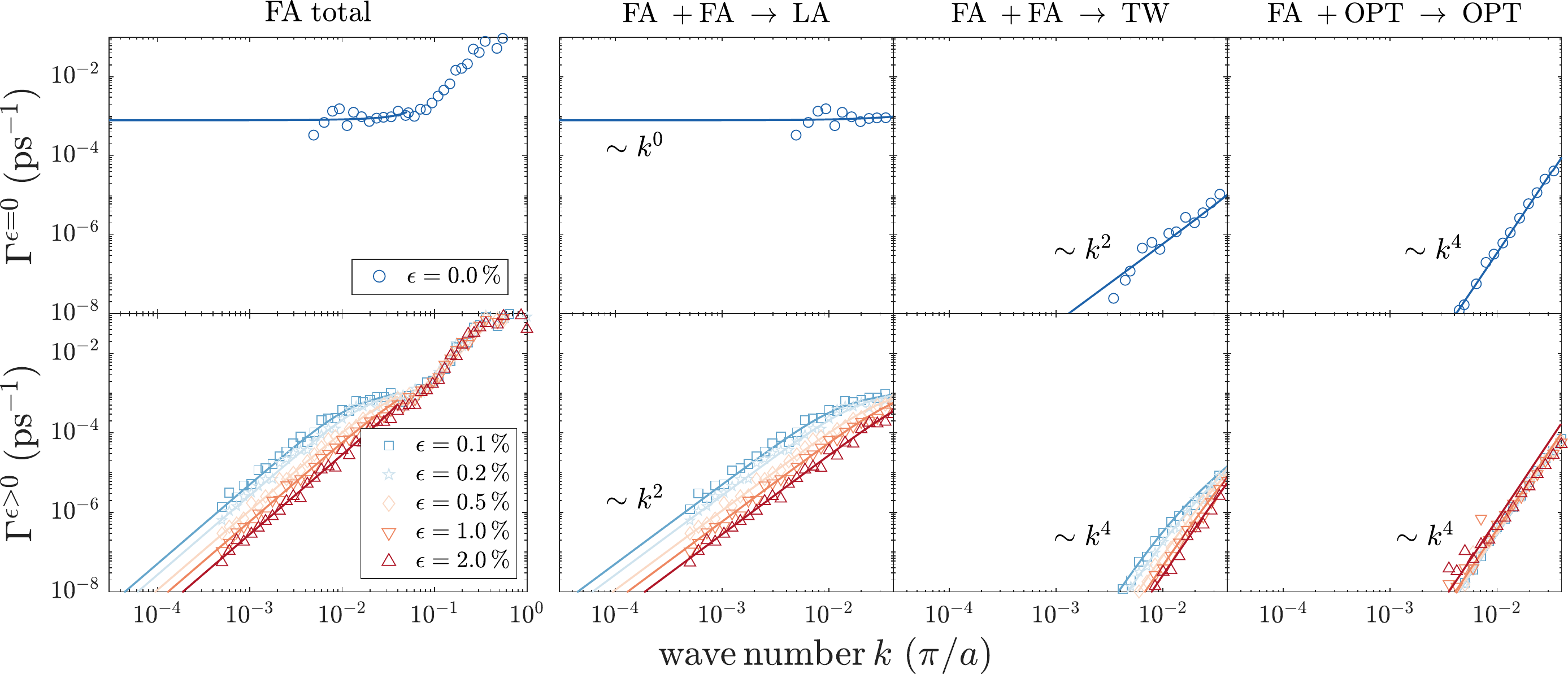}
  \caption{Same as Fig.~\ref{fig:2} but for FA modes. Both
    $\rm{FA}+\rm{FA} \rightarrow \rm{LA}$ and
    $\rm{FA}+\rm{FA} \rightarrow \rm{TW}$ absorption processes are
    affected by the strain induced frequency shift of FA modes in the
    limit $ka \ll 1$. As dictated by the dominant scattering channel
    $\rm{FA}+\rm{FA} \rightarrow \rm{LA}$, total FA scattering rates
    converge towards a constant in the case of stress-free tubes and
    gain a dependence $\sim \epsilon^{-1} k^2$ in the presence of
    strain.}
  \label{fig:3}
\end{figure*}

Generally speaking, the three-phonon scattering phase space for
long-wavelength FA modes is reduced in that no lower-lying phonon
branch providing pathways for kinematically allowed decay processes
exists. Our lattice-dynamical predictions of strain dependent FA
scattering rates are summarized in Fig.~\ref{fig:3}. In the limit
$ka \ll 1$, traceable scattering contributions arise from two types of
acoustic absorptions that depend critically on tensile strain, namely
$\rm{FA}+\rm{FA}\rightarrow \rm{LA}$ as well as
$\rm{FA}+\rm{FA}\rightarrow\rm{TW}$, and from optical interbranch
absorptions $\rm{FA}+\rm{OPT}\rightarrow\rm{OPT}$ which appear
unaffected by strain.

As is evident in Fig.~\ref{fig:3}, a reduction in the total scattering
rate of long-wavelength FA modes under increasing tensile lattice
strain derives from the dominant process
$\rm{FA}+\rm{FA}\rightarrow\rm{LA}$. The same three-phonon coupling
coefficient $V_{\rm{LFF}}$ enters into decay
$\Gamma_{\rm{LFF}}^{-}$ and absorption amplitudes
$\Gamma^{+}_{\rm{FFL}}$. Hence, according to our earlier
observation $V_{\rm{LFF}}\nsim \epsilon$ in the case of LA modes, a
reduction in the total rate should be solely ascribable to the strain
induced hardening of FA modes. Indeed, by setting
$V_{\rm{LFF}}=C_3kk'k''$ and by fitting our lattice-dynamical
predictions of $\Gamma^{+}_{\rm{FFL}}$ for different tensile strain
amplitudes $\epsilon$, we find no systematic variation of $C_{3}$ with
increasing tensile strain, leaving only $\omega_{\rm{FA}}$ as a strain
dependent factor. Based on analytical considerations of
$\Gamma^{+}_{\rm{FFL}}$, adopting either a quadratic
$\omega_{\rm{FA}}\sim k^2$, or a strictly linear FA mode dispersion
$\omega_{\rm{FA}}\sim \epsilon^{1/2}|k|$, stress-free and stretched tubes give
rise to $\Gamma_{\rm{FA}}^{\epsilon=0}\sim k^0$ and
$\Gamma_{\rm{FA}}^{\epsilon>0}\sim \epsilon^{-1}k^2$, respectively
(see Appendix \ref{sec:coal-two-flex}).

Comparing the absorption channels $\rm{FA}+\rm{FA}\rightarrow \rm{LA}$
and $\rm{FA}+\rm{FA}\rightarrow \rm{TW}$, the scaling relations of the
latter $\Gamma^{+}_{\rm{FFT}}\sim k^2$ in unstrained and
$\Gamma^{+}_{\rm{FFT}}\sim \epsilon^{-1}k^4$ in strained tubes stem
from $V_{\rm{TFF}}\sim kk'k''(k'-k'')$ (see Appendix
\ref{sec:coal-two-flex}). In our lattice-dynamical calculations of
$V_{\rm{TFF}}$, contributions that are cubic in the axial
wave number $k$ cancel out, in agreement with elasticity theory
predictions~\cite{MEG2009}. Similar cancellation effects manifest in
the case of optical absorption processes
$\rm{FA}+\rm{OPT}\rightarrow\rm{OPT}$. Following a standard
argument~\cite{Pitaevskii1981}, for a given long-wavelength acoustic
phonon $(j,k)$, one expects a linear scaling $V_{j\rm{OO}}\sim k$
to lowest order in $k$. In order to explain the scaling
$\Gamma^{+}_{\rm{FOO}}\sim k^4$ of optical interbranch
transitions in Fig.~\ref{fig:3}, however, we have to assume
$V_{\rm{FOO}}\sim k^2$ (see Appendix
\ref{sec:absorpt-optic-phon}).

\subsection{Twisting modes}
\label{sec:twisting-modes}

\begin{figure*}
  \includegraphics[width=\textwidth]{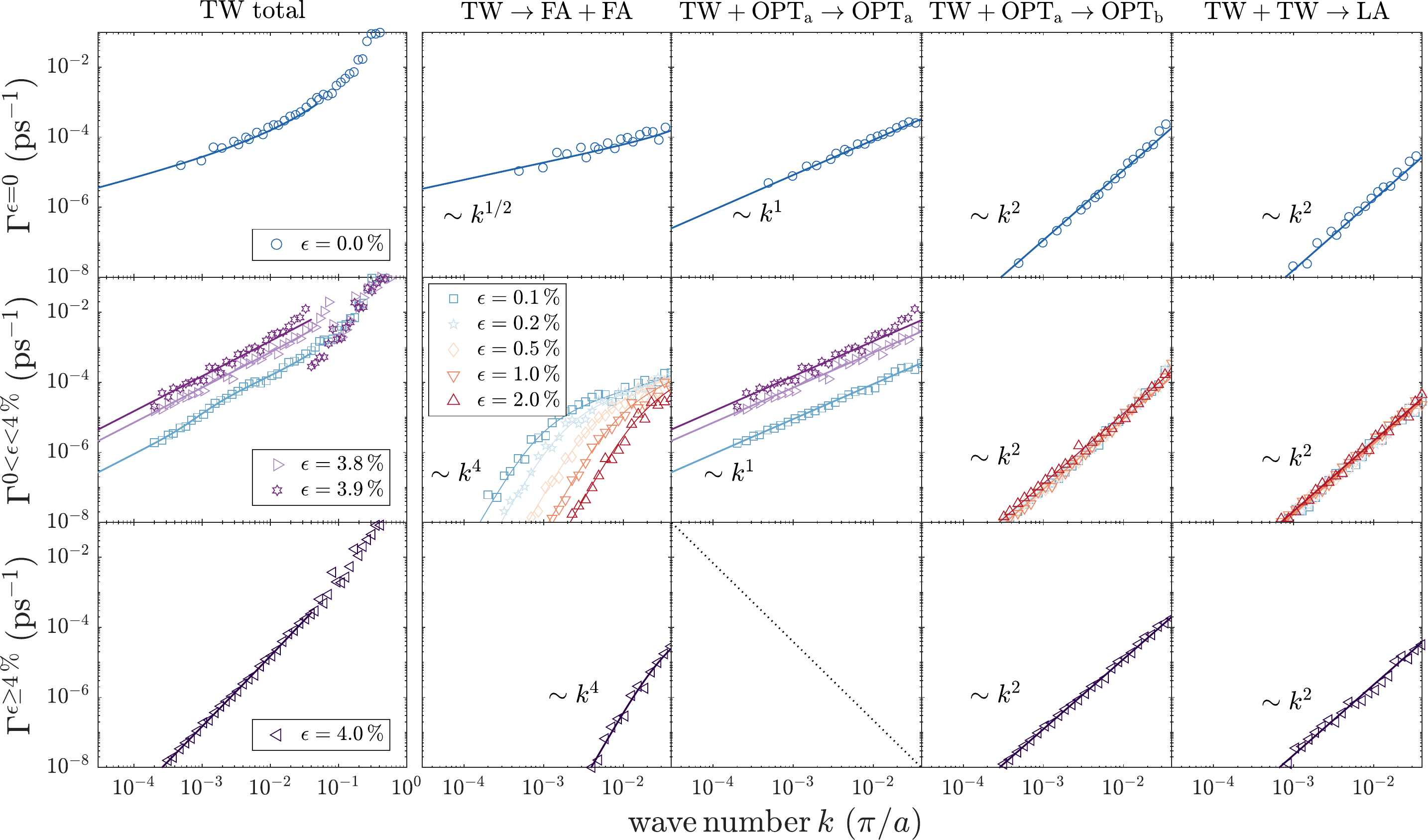}
  \caption{Same as Fig.~\ref{fig:2} but for TW modes. A third row of
    panels is added to distinguish the strain regimes
    $0<\epsilon<4\,\%$ (middle) and $\epsilon\geq 4\,\%$
    (bottom). Processes involving two optical modes are further
    subdivided into contributions resulting from intrabranch,
    $\rm{TW}+\rm{OPT}_a \rightarrow \rm{OPT}_a $ and interbranch,
    $\rm{TW}+\rm{OPT}_a\rightarrow\rm{OPT}_b$, transitions whose
    long-wavelength scaling is $\sim k$ and $\sim k^{2}$,
    respectively. For stress-free tubes, the decay process
    $\rm{TW}\rightarrow\rm{FA +\rm{FA}}$ goes as $\sim k^{1/2}$ and
    becomes the dominant contribution to the total rate in the limit
    $ka\ll 1$ . Under finite positive strain, $\epsilon >0$, this
    process develops a dependence $\sim \epsilon^{-7/2}k^4$ such that
    the scaling of the total becomes $\sim k$ due to optical
    intrabranch transitions, which are kinematically allowed as long
    as $\epsilon< 4\,\%$. In the strain regime $\epsilon \geq 4\,\%$,
    the total TW scattering rate inherits a dependence $\sim k^2$
    stemming from optical interbranch transitions.}
  \label{fig:4}
\end{figure*}

The branch of long-wavelength TW modes lies in the regime
$\omega_{\rm{FA}}<\omega_{\rm{TW}}<\omega_{\rm{LA}}$ and, as compared
to LA and FA modes, a kinematically less restricted three-phonon
scattering phase space may be anticipated. As shown in
Fig.~\ref{fig:4}, three-phonon transitions that contribute to the
dissipation of long-wavelength TW modes can be decomposed into two
types of low-energy acoustic and two types of high-energy optical
scattering channels. The former channels are
$\rm{TW}\rightarrow \rm{FA}+\rm{FA}$ and
$\rm{TW}+\rm{TW}\rightarrow \rm{LA}$, while the latter are given by
intra- and interbranch absorptions,
$\rm{TW}+\rm{OPT}_a\rightarrow \rm{OPT}_a$ and
$\rm{TW}+\rm{OPT}_a\rightarrow\rm{OPT}_b$, respectively. The relative
importance of these channels varies with tensile strain, causing a
nontrivial strain dependence in the total rate $\Gamma_{\rm{TW}}$.

In the case of unstrained tubes, $\epsilon=0$, the dominant scattering
amplitudes stem from the decay into two flexural modes,
$\Gamma^{-}_{\rm{TFF}}$, and from optical intrabranch absorptions,
$\Gamma^{+}_{\rm{TO_{a}O_{a}}}$, both of which are roughly of the
same order of magnitude within the resolved range of wave numbers. For
the decay $\rm{TW}\rightarrow \rm{FA}+\rm{FA}$, we have
$\Gamma^{-}_{\rm{TFF}}\sim k^{1/2}$ which is derivable given a
quartic scaling of $V_{\rm{TFF}}$ and provided that
$\omega_{\rm{FA}}\sim k^2$ under stress-free conditions (see Appendix
\ref{sec:decay-into-two}). Considering optical absorption processes of
acoustic modes, the scaling behavior of related scattering amplitudes
depends on whether momentum is transferred between two crossing
optical branches or within the same branch. If we adopt the
long-wavelength approximation
$V_{\rm{TOO}}\sim k$~\cite{Pitaevskii1981}, we find that
differences in the scattering phase space lead to
$\Gamma^{+}_{\rm{TO_{a}O_{a}}}\sim k$ and
$\Gamma^{+}_{\rm{TO_{a}O_{b}}}\sim k^2$, explaining the scaling
of intra- and interbranch transitions in Fig.~\ref{fig:4},
respectively (see Appendix \ref{sec:absorpt-optic-phon}).

In the presence of tensile strain, $\epsilon>0$, as before in the case
of $\Gamma^{-}_{\rm{LFF}}$, $\Gamma^{+}_{\rm{FFL}}$, and
$\Gamma^{+}_{\rm{FFT}}$, frequency hardening of long-wavelength FA
modes strongly suppresses the scattering amplitude of
$\rm{TW}\rightarrow\rm{FA}+\rm{FA}$ decay. Applying a fit to our
lattice-dynamical predictions of $\Gamma^{-}_{\rm{TFF}}$, we again
notice that strain has no influence on the coupling coefficient
$V_{\rm{TFF}}$, whose quartic wave number dependence together with
$\omega_{\rm{FA}}\sim \epsilon^{1/2}|k|$ yields
$\Gamma^{-}_{\rm{TFF}}\sim \epsilon^{-7/2}k^4$ (see Appendix
\ref{sec:decay-into-two}). For moderately strained tubes, total TW
scattering rates in the long-wavelength limit remain lower bounded by
optical intrabranch transitions and the limiting behavior transitions
from $\sim k^{1/2}$ to $\sim k$.

\begin{figure}
  \includegraphics[width=0.9\columnwidth]{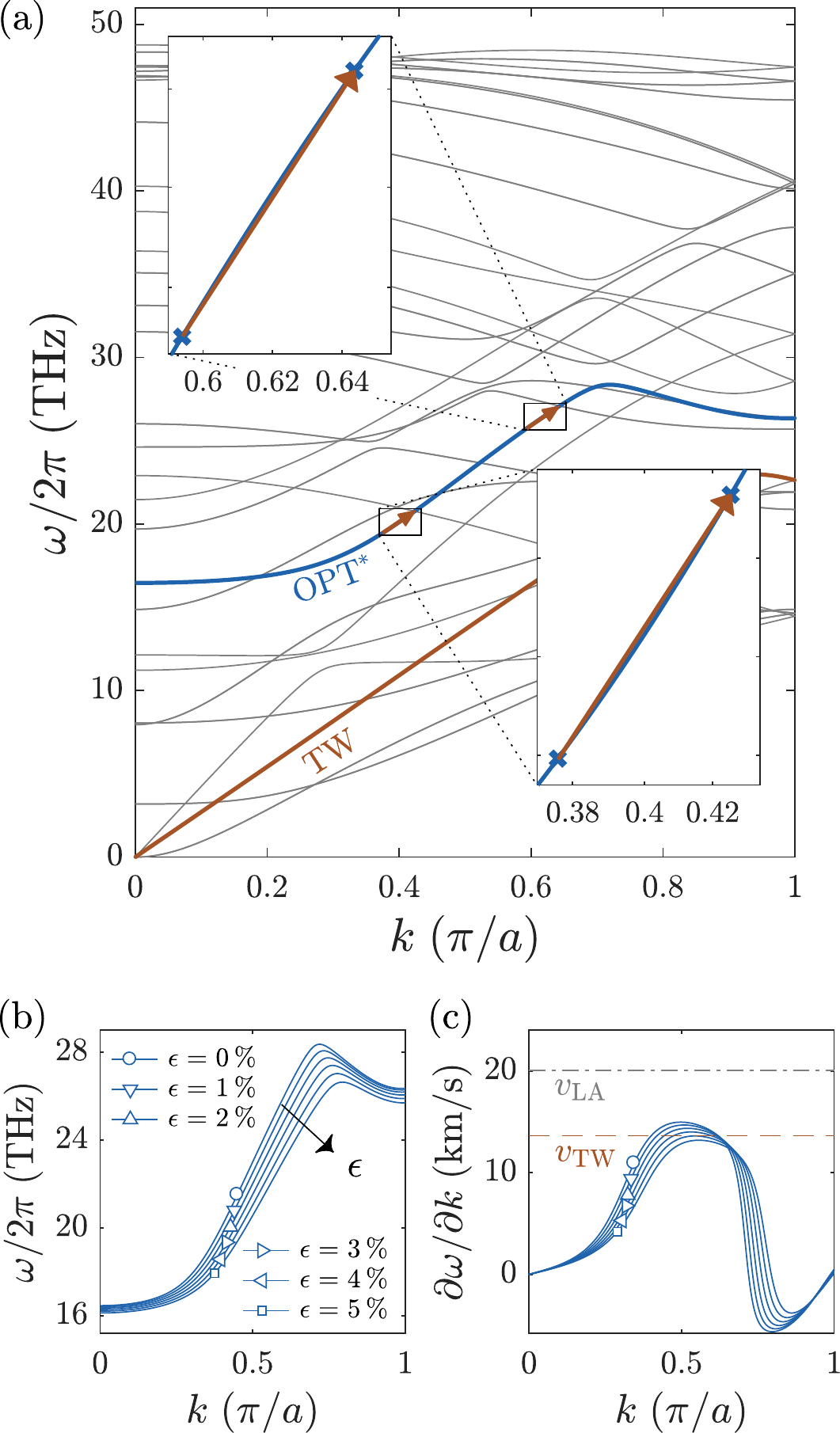}
  \caption{(a) The spectrum of harmonic phonon modes of a (4,4) carbon
    nanotube at $\epsilon=0$ comprises one optical branch
    ($\rm{OPT}^*$) whose slope matches the slope of the TW branch,
    enabling intrabranch transitions
    $\rm{TW}+\rm{OPT}_a\rightarrow \rm{OPT}_a$. (b) Frequency
    softening of the $\rm{OPT}^*$ branch under tensile strain. (c) The
    slope of the $\rm{OPT}^*$ branch decreases with increasing strain
    amplitude. At $\epsilon=4\,\%$, the velocity $v_{\rm{TW}}$ exceeds
    $\partial \omega_{\rm{OPT}^*}/\partial k$ and the scattering of
    long-wavelength TW modes due to intrabranch transitions is
    kinematically prohibited.}
  \label{fig:5}
\end{figure}

Interestingly, at larger strain amplitudes, long-wavelength scattering
contributions arising from optical intrabranch absorption processes
may become susceptible to lattice strain as well. For our model of a
(4,4) carbon nanotube, we find that optical intrabranch transitions
can occur only in the regime $\epsilon<4\,\%$ beyond which the total
TW rate abruptly gains a $k^2$ dependence resulting from optical
interbranch transitions. While no structural changes can be observed
in the carbon lattice at $\epsilon=4\,\%$, an explanation for the
discontinuity in the total TW rate is found by considering the strain
dependence of optical phonon frequencies, see Fig.~\ref{fig:5}. A
necessary condition for the realization of intrabranch transitions is
the existence of a supersonic optical branch $\omega_{\rm{OPT^*}}$. As
illustrated in Fig.~\ref{fig:5}(a), one such branch whose slope
$\partial \omega_{\rm{OPT}^*}/\partial k$ surpasses the slope of the
TW branch can be singled out in the spectrum of high-energy optical
modes. Under tensile loading, the branch $\omega_{\rm{OPT}^*}$
softens, cf. Fig.~\ref{fig:5}(b), and its slope
decreases. Figure~\ref{fig:5}(c) demonstrates that for strain
amplitudes $\epsilon\geq 4\,\%$ the necessary condition
$\partial \omega_{\rm{OPT}^*}/\partial k>v_{\rm{TW}}$ for optical
intrabranch transitions is no longer fulfilled. A somewhat
pathological prediction of Eq.~\eqref{eq:2} reveals itself as the
critical strain amplitude $\epsilon=4\,\%$ is approached from
below. Here, scattering rates are predicted to increase, which can be
traced to the vanishing curvature,
$\partial^2\omega_{\rm{OPT}^*}/\partial k^2\rightarrow 0$, in the wave
number regions depicted by the insets of Fig.~\ref{fig:5}(a), cf.
Eqs.~\eqref{eq:7} and \eqref{eq:25}. In passing, we can compare TW and
LA mode velocities in Fig.~\ref{fig:5}(c) to conclude that
long-wavelength LA modes remain protected from optical intrabranch
transitions.

We point out that a recent numerical study on unstrained nanowires by
Rashid et al.~\cite{Rashid2018} found low-frequency power-law
dependencies of three-phonon scattering rates for LA, FA, and TW modes
that conform to our results for unstrained tubes. From the viewpoint
of acoustic phonons, nanowires and nanotubes are very similar. It
seems therefore natural to conjecture that considerations in the
present study apply to a broader class of one-dimensional materials,
implying, for example, that a strain induced reduction in the
scattering strength of long-wavelength acoustic phonons can be
achieved in nanowires as well.

\section{Thermal conductivity \\under the relaxation time
  approximation}
\label{sec:therm-cond-}

\begin{figure}
  \includegraphics[width=0.85\columnwidth]{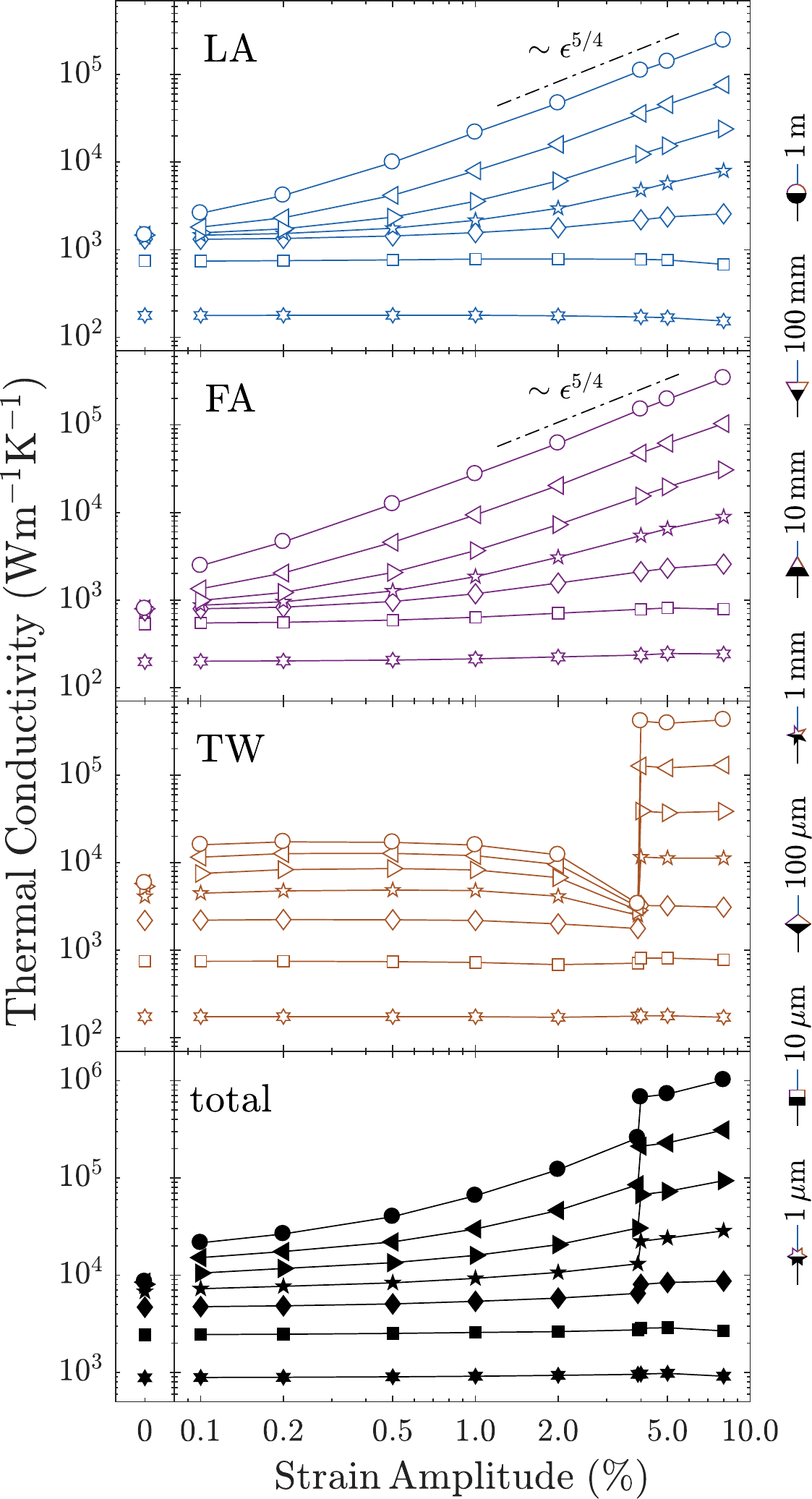}
  \caption{Lattice thermal conductivity under the relaxation time
    approximation of a pristine (4,4) carbon nanotube at room
    temperature as derived from strain-dependent three-phonon
    scattering rates. In stress-free tubes, $\epsilon=0$, the
    conductivity converges with tube length $L$. Under finite tensile
    strain, $\epsilon >0$, and in the limit of long tubes, the
    acoustic phonon contributions to the lattice thermal conductivity
    scale as $\kappa_{\rm{LA}/\rm{FA}}\sim \epsilon^{5/4}L^{1/2}$ and
    as $\kappa_{\rm{TW}}\sim \ln{L}\;(\sim L^{1/2})$ for
    $\epsilon<4\,\%\;(\geq 4\,\%)$.}
  \label{fig:6}
\end{figure}

Three-phonon scattering rates presented in
Figs.~\ref{fig:2}-\ref{fig:4} are a key ingredient for determining the
lattice thermal conductivity of pristine nanotubes in the framework of
the PBE. Under the single-mode relaxation time approximation
(RTA)\cite{srivastava1990physics}, the thermal conductivity of carbon
nanotubes is given by
\begin{align}
  \label{eq:5}
  \kappa^{\rm{RTA}}=\frac{1}{A}\sum_{j}\int^{\pi/a}_{-\pi/a}\frac{{\rm{d}}k}{2\pi}\,\hbar
  \omega_{\nu}\frac {\partial n_{\nu}}{\partial
  T}v_{\nu}^2\tau_{\nu},
\end{align}
where typically an annular cross-sectional area $A=\pi D \delta$ is
adopted with thickness $\delta=\SI{3.35}{\angstrom}$ corresponding to
the interlayer distance in graphite, the phonon velocity is taken as
the slope of dispersion branches,
$v_{\nu}=\partial \omega_{v}/\partial k$, and $\tau_{\nu}$ denotes the
phonon relaxation time. Following the standard practice within the PBE
formalism, intrinsic three-phonon scattering rates as per
Eqs.~\eqref{eq:1} and~\eqref{eq:2} are combined with a boundary scattering
rate to compute phonon relaxation times in tubes of finite length $L$,
\begin{align}
  \label{eq:6}
  \frac{1}{\tau_{\nu}}=\Gamma_{\nu}+\frac{2|v_{\nu}|}{L}.
\end{align}
The boundary scattering term acts as an upper bound to phonon
relaxation times, which effectively filters out contributions from
low-frequency modes with small intrinsic scattering rates and ensures
a ballistic transport regime $\kappa\sim L$ in the limit of short
tubes~\cite{MB05}. Given the scaling laws of anharmonic phonon
scattering $\Gamma_{j}(k)\sim \epsilon^{r}k^{s}$ found in
Sec.~\ref{sec:mode-level-analysis}, the integration over acoustic
branch contributions in Eq.~\eqref{eq:5} converges for
${L\rightarrow\infty}$ in stress-free tubes, $\epsilon=0$, but
diverges in the presence of strain, $\epsilon >0$ (see Appendix
\ref{sec:asympt-stra-tube}).

Usually, per-mode conductivities as per Eq.~\eqref{eq:5} are sampled
and summed over a finite grid of wave numbers,
$\kappa^{\Sigma}=N^{-1}\sum_j\sum_{k}\kappa_j(k)$, with grid spacing
$\Delta k=2\pi/N a$, which requires to test convergence with respect
to $N$. Such an approach, however, is problematic if intrinsic
lifetimes exhibit a divergence,
$(\Gamma_{\nu})^{-1}\rightarrow \infty$ for $|k|\rightarrow 0$, as is
the case for LA and FA modes in strained and for TW modes in both
unstrained and strained tubes. For large but finite values of $L$,
sufficiently dense wave number grids to converge the thermal
conductivity might be out of reach by means of lattice-dynamical
calculations. Here, for our model of a (4,4) carbon nanotube, we use
the asymptotic fits to dispersion laws and three-phonon scattering
rates as depicted in Figs.~\ref{fig:1}-\ref{fig:4} in order to
extrapolate the wave number dependencies of acoustic per-branch
conductivities in the ${|k|\rightarrow 0}$ limit. This in turn allows
us to numerically integrate acoustic long-wavelength contributions all
the way down to the Gamma point $(k=0)$,
$\kappa^{\rm{LW}}=\int_{0}{\rm{d}}k\,\kappa_j(k)$ for
$j\in \{\rm{LA},\rm{FA},\rm{TW}\}$.

\begin{figure}[h!]
  \includegraphics[width=0.8\columnwidth]{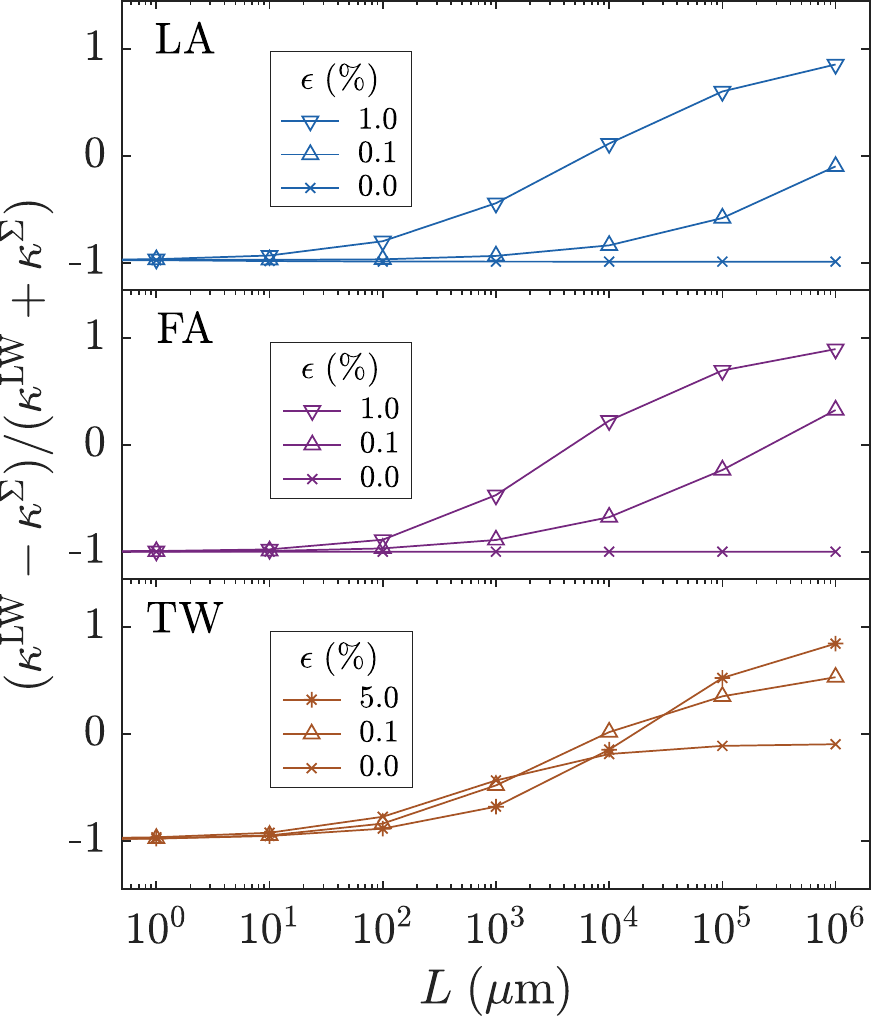}
  \caption{Relative importance of summed $\kappa^{\Sigma}$
    vs integrated $\kappa^{\rm{LW}}$ acoustic thermal conductivity
    contributions as a function of tube length $L$ and strain
    amplitude $\epsilon$. $\kappa^{\Sigma}$ corresponds to a summation
    over a wave number grid with $N=1000$ points. Extrapolated
    contributions stemming from long-wavelength acoustic modes in
    close proximity to the Gamma point $(k=0)$ are captured by
    $\kappa^{\rm{LW}}$ but not included in $\kappa^{\Sigma}$. A value
    of $-1$ on the vertical axis implies a convergence of thermal
    conductivity with respect to the wave number grid resolution $N$,
    whereas larger values justify an extrapolation approach.}
  \label{fig:7}
\end{figure}

In Fig.~\ref{fig:6}, we combine per-mode conductivities sampled over a
grid with $N=1000$ points with acoustic mode contributions integrated
over the interval $[-2\pi/Na;\,2\pi/Na]$,
$\kappa=\kappa^{\Sigma}+\kappa^{\rm{LW}}$, to show the conductivity
predictions according to Eqs.~\eqref{eq:5} and~\eqref{eq:6} in the
limit of small strain amplitudes $\epsilon$. The relative importance
of summed $\kappa^{\Sigma}$ vs integrated long-wavelength
contributions $\kappa^{\rm{LW}}$ is displayed in Fig.~\ref{fig:7}. As
can be seen in Fig.~\ref{fig:6}, relatively long tubes
$L> \SI{100}{\mu m}$ are required to observe an enhancement of thermal
transport due to a reduction in the scattering rates of
long-wavelength acoustic modes in the presence of strain. For a tube
with fixed but very large $L$, frequency resolved conductivity
contributions shift towards lower frequencies with increasing strain
amplitude. The lowest contributing frequencies, however, are
ultimately determined by the tube length since, in principle,
low-frequency acoustic modes of wave number $k$ cannot exist in tubes
of size $L$ unless $1/k < L$.

In the long-tube limit, LA modes in strained tubes yield a
conductivity contribution that scales as
${\sim \epsilon^{5/4}L^{1/2}}$ which follows from
$\Gamma_{\rm{LA}}\sim \epsilon^{5/2} k^2$. The same scaling relation
is obtained in the case of FA branch contributions which, in this
case, is attributable to ${\Gamma_{\rm{FA}}\sim \epsilon^{-1}k^{2}}$
as well as $\omega_{\rm{FA}}\sim \epsilon^{1/2}k$. For TW modes, the
long-wavelength scaling relation $\Gamma_{\rm{TW}}\sim k$ leads to a
logarithmically divergent conductivity contribution in tubes subject to
strain amplitudes $0<\epsilon< 4\,\%$. The discontinuity in the strain
dependence of long-wavelength TW scattering rates at $\epsilon=4\,\%$,
as detailed in Fig.~\ref{fig:5}, is filtered out in short tubes but
becomes apparent if $L>\SI{10}{\mu m}$. For strain amplitudes
$\epsilon\geq 4\,\%$, a divergence $\sim L^{1/2}$ of the TW branch
conductivity contribution is implied by $\Gamma_{\rm{TW}}\sim k^2$. In
the bottom panel of Fig.~\ref{fig:6}, we show total thermal
conductivity predictions including optical branch contributions which,
however, do not increase significantly with tube length, and are
therefore of minor importance in the large-$L$ regime. Remarkably,
under the RTA, long-wavelength TW modes are predicted to be the
dominant heat carriers in unstrained tubes as the tube length
approaches the millimeter length scale.

As can be inferred from Fig.~\ref{fig:7}, per-branch acoustic
conductivity data sampled over a grid with $N=1000$ discrete wave
numbers gives a poor representation of the total thermal conductivity
according to Eqs.~\eqref{eq:5} and~\eqref{eq:6} as the tube length $L$
becomes large. In unstrained tubes, $\epsilon=0$, it is the square
root divergence of intrinsic TW lifetimes
$(\Gamma_{\rm{TW}})^{-1}\sim k^{-1/2}$ which requires grid
calculations with $N\gg 1000$ to converge conductivity predictions in
the long-tube limit. Convergence issues with respect to the wave
number grid resolution $N$ are generally exacerbated in the presence
of tensile strain, $\epsilon>0$, where all three acoustic mode
polarizations exhibit a singularity of the form
$(\Gamma_{j})^{-1} \sim k^{-r}$ with $r\geq 1$ in the long-wavelength
limit.

\section{Discussion}
\label{sec:discussion}

Two approximations have been employed in the present study that
deserve further scrutiny in the future.  For one, phonon frequency
shifts at finite temperature induced by lattice anharmonicity have
been assumed to be negligible. In the case of two-dimensional
crystals, by taking into account the real part of the phonon
self-energy, it was noted~\cite{Michel2015} that out-of-plane
vibrations are renormalized at finite temperature to the extent that
the quadratic dispersion in stress-free samples under the harmonic
lattice approximation becomes linear in the long-wavelength limit. For
unstrained tubes, the obtained scaling laws of three-phonon scattering
presented in Sec.~\ref{sec:mode-level-analysis} critically depend on
the condition that FA mode frequencies vanish quadratically in the
long-wavelength limit, $\omega_{\rm{FA}} \sim k^2$. Taking a
temperature dependent, possibly linearized, FA mode dispersion as
input to a perturbative calculation of phonon-phonon interactions,
low-frequency dissipation rates $\Gamma_{j}(|k|\rightarrow 0)$ could
be drastically decreased, as we demonstrate in this work by examining
tubes under tensile load. In a similar context, while studying
phonon-phonon interactions in a semiflexible monoatomic chain,
Santhosh and Kumar~\cite{Santhosh2010} pointed out the necessity of
anharmonic frequency renormalization, since the standard perturbative
treatment of harmonic phonon modes violates the notion of well-defined
quasiparticles, requiring $\Gamma_{j}(k) < \omega_{j}(k)$ as
$|k|\rightarrow 0$. In particular, according to
Ref.~\cite{Santhosh2010}, longitudinal, $\omega_x\sim k$, and
transverse, $\omega_{y}\sim k^2$, modes of a semiflexible chain under
the harmonic lattice approximation give rise to
$\Gamma_x\sim |k|^{-1/2}$ and $\Gamma_{y}\sim k^0$, in agreement with
our findings for LA and FA modes in unstrained tubes,
cf. Sec.~\ref{sec:longitudinal-modes} and
Sec.~\ref{sec:flexural-modes}. Effective long-wavelength dissipation
rates under phonon renormalization, as predicted
in Ref.~\cite{Santhosh2010}, are instead significantly reduced.

Another point worthy of future research relates to the approximation
of thermal conductivity. Previous studies emphasized the inadequacy of
the RTA in the case of
one-~\cite{LBM09,Wang2017a,Pandey2018,Barbalinardo21} and
two-dimensional~\cite{Lindsay2010b,Fugallo2014,Cepellotti2015,Cepellotti2016}
materials, noting the importance of obtaining the thermal conductivity
from an exact solution of the PBE. Going beyond the RTA, one is
confronted with numerically solving a large set of linear equations~
\cite{srivastava1990physics,Chernatynskiy2010},
\begin{align}
  \label{eq:8} X_{\nu}=\sum_{\nu'}P_{\nu \nu'}\psi_{\nu'}.
\end{align}
Here, the inhomogeneity $X_{\nu}$ and the collision kernel
$P_{\nu\nu'}$ are given in terms of phonon dispersion data and phonon
scattering rates, respectively, while the unknowns $\psi_{\nu}$ are
the nonequilibrium phonon deviation functions, which ultimately
determine the thermal conductivity,
$\kappa \sim N^{-1}\sum_{\nu}X_{\nu}\psi_{\nu}$ (up to some
prefactors)~\cite{srivastava1990physics,Chernatynskiy2010}. The RTA
approach to lattice heat transport, as pursued in
Sec.~\ref{sec:therm-cond-}, is tantamount to keeping only the diagonal
terms of the collision kernel,
$(P\psi)_{\nu}\approx
P_{\nu}\psi_{\nu}$~\cite{srivastava1990physics,Chernatynskiy2010}. It
is sometimes noted that RTA predictions generally underestimate the
true thermal conductivity~\cite{Allen2013,Ma2014}. Therefore, given
our results presented in Sec.~\ref{sec:therm-cond-}, one is tempted to
conclude that nanotubes under tensile strain act as superdiffusive
heat conductors which give rise to $\kappa \sim L^\eta$ with
$ \eta\geq 1/2$ in the long-tube limit. In the case of graphene, a
similar reasoning based on the RTA was applied by the authors of
Ref.~\cite{BGM2012} to conclude that the thermal conductivity of
isotropically strained sheets should be logarithmically divergent with
domain size. In pursuing an exact solution approach, however, some of
the same authors later arrived at the conclusion that the conductivity
of pristine graphene should remain upper bounded, even in the presence
of strain~\cite{Fugallo2014}. By taking into account the low-frequency
three-phonon scattering rates of Sec.~\ref{sec:mode-level-analysis},
it would be desirable to make more rigorous analytical statements
about the long-tube limit of thermal conductivity based on
Eq.~\eqref{eq:8}. Very recently, a convergence with tube length was
reported by Barbalinardo et al.~\cite{Barbalinardo21}, who derived the
thermal conductivity of a (10,0) nanotube in stress-free condition by
numerically inverting the full collision kernel $P_{\nu\nu'}$. Under
the assumption of tubes in perfect mechanical equilibrium, such
convergent behavior falls into line with RTA level predictions. Still,
the role of tensile strain within exact solution approaches to the PBE
remains to be clarified.

\section{Conclusions}
\label{sec:conclusion}

Taking as example a (4,4) carbon nanotube, we have performed numerical
lattice-dynamical calculations to address the role of acoustic phonons
in low-dimensional nanotube heat transport. By supporting our
calculations with analytical considerations and by making recourse to
continuum theories, we derived the general long-wavelength scaling
relations of acoustic phonon scattering rates that follow from
standard anharmonic perturbation theory. Based on our model, the onset
of long-wavelength behavior in tubes of radius $R$ is predicted at
around $|k|R<0.3$. We expect that somewhat smaller wave numbers are
needed in the case of chiral nanotubes, which tend to have larger
translational unit cells.

As was shown earlier in the case of graphene~\cite{BGM2012}, the
three-phonon scattering strength of long-wavelength acoustic modes can
be reduced by tensile lattice strain whenever acoustic-flexural mode
type processes are dominant. The scattering phase space for these
processes is greatly reduced by a strain-induced hardening of flexural
mode frequencies. In going from a two-dimensional phonon scattering
phase space in graphene to a one-dimensional one, this effect becomes
even more pronounced in nanotubes. As compared to longitudinal modes,
long-wavelength twisting modes in nanotubes couple only weakly to
flexural modes. Twisting modes have divergent anharmonic lifetimes in
the long-wavelength limit in both unstrained and strained tube
configurations, which makes $k$-space integration in the PBE-RTA
formalism challenging.

Arguably, for nanotubes under tensile strain, PBE-RTA predictions
provide strong evidence for the nonexistence of an upper bound to
thermal conductivity, suggesting the possibility of superdiffusive
heat transport in macroscopically long nanotubes. Whether exact
solution approaches to the PBE, or the systematic inclusion of
four-phonon scattering processes, help to renormalize a formally
divergent thermal conductivity awaits further investigation.

\begin{acknowledgements}

  The authors gratefully acknowledge financial support from the
  Discovery Grant Program of the Natural Sciences and Engineering
  Research Council of Canada. D.B.~was supported by the QuEST
  fellowship at the University of British Columbia. This research was
  undertaken thanks, in part, to funding from the Canada First
  Research Excellence Fund, Quantum Materials and Future Technologies
  Program.
  
\end{acknowledgements}

\clearpage
\newpage

\onecolumngrid
\appendix

\section{Three-phonon coupling coefficients}
\label{sec:three-phon-coupl}

For a one-dimensional monoatomic crystal, the three-phonon coupling
coefficients entering into Eq.~\eqref{eq:2} are given
by~\cite{Born1954,Z60}
\begin{align}
  \label{eq:59}    
  V_{jj'j''}(k,k',k'')=\frac{1}{m^{3/2}}\sum_{L',L''}e^{ik'r_{L'}}e^{ik''r_{L''}}
  \,\sum_{l,l',l''}\sum_{\alpha,\alpha',\alpha''}w_{l}^{\alpha}(j,k)w_{l'}^{\alpha'}(j',
  k')w_{l''}^{\alpha''}(j'',k'')\Phi_{0l,L'l',L''l''}^{\alpha,\alpha',\alpha''},
\end{align}
where $m$ denotes the atomic mass, the first sum runs over
translational unit cells residing at $r_{L'}\,(r_{L''})$, the second
sum runs over atomic sites within a unit cell, the third sum extends
over Cartesian coordinates, the $w$'s are the harmonic phonon
eigenvectors, and $\Phi$ is the third order force constant tensor.

For carbon nanotubes, in order to accurately determine the coupling
coefficients of acoustic phonons in the limit of small wave numbers, it
is crucial to ensure that the tensor $\Phi$ obeys translational and
rotational sum rules~\cite{Popov2000}. By adopting an empirical
Tersoff type interaction potential and by deriving force constants
analytically, these sum rules are naturally captured in our
calculations.

\section{Three-phonon scattering rates in the low-frequency limit}
\label{sec:asympt-scal-laws}

Starting from Eqs.~\eqref{eq:1} and~\eqref{eq:2}, we aim to determine the
scaling of the transition amplitudes $\Gamma^{\pm}_{jj'j''}(k)$ for
acoustic modes, $j\in \{\rm{LA},\rm{FA},\rm{TW}\}$, in the
long-wavelength limit $0<ka \ll 1$. The quasimomentum selection rule
$\Delta_{k\pm k' -k'',2\pi m/a}$ in Eq.~\eqref{eq:2} can be invoked to
resolve the sum over $k''$ in Eq.~\eqref{eq:1}. Setting
$k''=k\pm k'-2\pi m/a$, the integer $m=0,\pm1$ is chosen such that the
sum $k\pm k'$ lies within the first zone $[-\pi/a;
\pi/a]$. Three-phonon scattering events involving low-frequency
phonons necessarily correspond to normal processes
$(m=0)$~\cite{Han1993}, and for a given mode triplet $(j,j',j'')$ we
have
\begin{align}
  \label{eq:777}
  \begin{split}
    \Gamma^{\pm}_{jj'j''}(k)=\frac{a}{2\pi}\int^{\pi/a}_{-\pi/a}{\rm{d}}k'\,&\frac{\hbar
      \pi}{4\omega_{j}(k) \omega_{j'}(k') \omega_{j''}(k\pm
      k')}\frac{\left[n_{j'}(k')+\frac{1}{2}\pm\frac{1}{2}\right]n_{j''}(k\pm
      k')}{n_{j}(k)}\\
    &\times|V_{jj'j''}(k,\pm k',-k\mp
    k')|^2\,\delta(\Omega^{\pm}_{jj'j''}(k,k') ),
  \end{split}
\end{align}
where the function
$\Omega^{\pm}_{jj'j''}(k,k')=\omega_{j}(k)\pm\omega_{j'}(k')-\omega_{j''}(k\pm
k')$ is introduced to represent the condition of energy
conservation. Noting that
$n_j(k)=[\exp(\hbar \omega_{j}(k)/k_{\rm{B}}T)-1]^{-1}\approx
k_{{\rm{B}}}T/\hbar \omega_{j}(k)$, Eq.~\eqref{eq:777} becomes
\begin{align}
  \label{eq:7}
  \Gamma^{\pm}_{jj'j''}(k)\sim
  \frac{\left[n_{j'}(k^{*})+\frac{1}{2}\pm\frac{1}{2}\right]n_{j''}(k\pm
  k^{*})}{\omega_{j'}(k^*)\
  \omega_{j''}(k\pm k^*)}\frac{|V_{jj'j''}(k,\pm k^*,-k\mp
  k^*)|^2}{\left| \frac{\partial \Omega^{\pm}_{jj'j''}(k,k^*)}{\partial k'}\right|},
\end{align}
where the wave number $k^*=k^*_{jj'j''}(k)$ follows from
$\Omega^{\pm}_{jj'j''}(k,k^*)=0$.

\subsection{Decay into two flexural modes}
\label{sec:decay-into-two}

Let us first consider transitions where a long-wavelength acoustic
phonon from a linear branch decays into two flexural acoustic modes,
$\rm{LA/TW}\rightarrow \rm{FA}+\rm{FA}$. As per
Eqs.~\eqref{eq:3} and~\eqref{eq:4}, $k^*$ is the positive root of
\begin{align}
  \label{eq:245}
  \Omega^{-}(k,k^*)=v_{j}k-\sqrt{\mathcal{T}_1 \epsilon
  {k^*}^2+\mathcal{Y}(\epsilon){k^*}^4}-\sqrt{\mathcal{T}_1\epsilon \left(k-k^*\right)^2+\mathcal{Y}(\epsilon)\left(k-k^*\right)^4}=0,
\end{align}
where $v_{j}=v_{\rm{LA}/\rm{TW}}$ and
$\mathcal{Y}(\epsilon)=v^2_{\rm{LA}}R^2/2-\mathcal{T}_2\epsilon$. The
condition of energy conservation can be resolved analytically in two
limiting cases by either assuming a strictly quadratic dispersion of
flexural modes under stress-free conditions,
$\omega_{\rm{FA}}\sim k^2$, or by adopting a strictly linear
dispersion in the presence of tensile strain $\epsilon$,
$\omega_{\rm{FA}}\sim \epsilon^{1/2}k$. In the former scenario, we
have
\begin{align}
  \label{eq:9}
  k^*=\frac{k}{2}+\sqrt{\frac{v_{j}}{2\,\mathcal{Y}(0)}k-\frac{k^2}{4}}\quad\text{and}\quad\left|\frac{\partial
  \Omega^{-}(k,k^*)}{\partial k'}\right|=2\sqrt{2\sqrt{\mathcal{Y}(0)}v_{j}k-\mathcal{Y}(0)k^2},
\end{align}
where both expressions go as $k^{1/2}$. In the low-frequency regime,
$n_{\rm{FA}}(k)\approx k_{\rm{B}}T/\hbar\omega_{\rm{FA}}(k)$, and by
combining Eqs.~\eqref{eq:7} and~\eqref{eq:9} we find
\begin{align}
  \label{eq:88}
  \Gamma^{-}_{j{\rm{F}\rm{F}}}(k)\sim k^{-9/2}
  \left|V_{j{\rm{FF}}}\right|^2 \sim
  \begin{dcases}k^{-1/2}&
    \text{if}\; j=\rm{LA},\\
    k^{1/2}&
    \text{if}\; j=\rm{TW},
  \end{dcases}
\end{align}
where the standard long-wavelength
approximation~\cite{Z60,Pitaevskii1981} $V_{\rm{LFF}}\sim kk'k''$
applies to $\rm{LA}\rightarrow \rm{FA}+\rm{FA}$ but cancellation
effects in Eq.~\eqref{eq:59} lead to
$V_{\rm{TFF}}\sim kk'k''(k'-k'')$~\cite{MEG2009} in the case of
$\rm{TW}\rightarrow\rm{FA}+\rm{FA}$.

If, on the other hand, a strictly linear flexural mode dispersion is
assumed, $\omega_{\rm{FA}}\sim \epsilon^{1/2}k$, we have
\begin{align}
  \label{eq:555}
  k^*=\left(1+\frac{v_{j}}{\sqrt{\mathcal{T}_1\epsilon}}\right)\frac{k}{2}\quad\text{and}\quad\left|\frac{\partial
  \Omega^{-}(k,k^*)}{\partial k'}\right|=2\sqrt{\mathcal{T}_1\epsilon},
\end{align}
which, when plugged into Eq.~\eqref{eq:7}, leads to
\begin{align}
  \label{eq:99}
  \Gamma^{-}_{j{\rm{F}\rm{F}}}(k)\sim \epsilon^{-5/2}k^{-4}
  \frac{\left|V_{j{\rm{FF}}}\right|^{2}}{\left(1-v^2_{j}/\mathcal{T}_1\epsilon\right)^{2}} \sim
  \begin{dcases}\epsilon^{-5/2}k^{2}&
    \text{if}\; j=\rm{LA},\\
    \epsilon^{-7/2}k^{4}& \text{if}\; j=\rm{TW},
  \end{dcases}
\end{align}
where the long-wavelength approximations of
$V_{\rm{LA}\rm{FA}\rm{FA}}$ and $V_{\rm{TW}\rm{FA}\rm{FA}}$ remain
unaffected by tensile lattice strain, as suggested by our numerical
evaluation of Eq.~\eqref{eq:59}.

\subsection{Coalescence of two flexural modes}
\label{sec:coal-two-flex}

To determine the scaling of $\Gamma^{+}_{{{\rm{FA}\rm{FA}}}j''}$ for
$j''\in \{\rm{LA},\rm{TW}\}$, we again solve the condition of energy
conservation $\Omega^{+}(k,k^*)=0$ for $k^*$ and consider the cases
$\omega_{\rm{FA}}\sim k^2$ and
$\omega_{\rm{FA}}\sim \epsilon^{1/2}k^2$ separately. In the unstrained
case, $\epsilon=0$, we have
\begin{align}
  \label{eq:67}
  k^*=\frac{v_{j''}}{2\sqrt{\mathcal{Y}(0)}} -\sqrt{\frac{v_{j''}^2}{4\mathcal{Y}(0)}+\frac{v_{j''}}{\sqrt{\mathcal{Y}(0)}}k-k^2}\quad\text{and}\quad\left|\frac{\partial
  \Omega^{+}(k,k^*)}{\partial k'}\right|=\sqrt{v_{j''}^2+4\sqrt{\mathcal{Y}(0)}v_{j''}\,k-4\mathcal{Y}(0)k^2}.
\end{align}
Here, the first expression goes as $k$ and the second converges
towards $v_{j''}$. In evaluating Eq.~\eqref{eq:7}, we can assume
${n_{\rm{FA}}(k)+1\approx k_{\rm{B}}T/\hbar \omega_{\rm{FA}}(k)}$ and
find
\begin{align}
  \label{eq:61}
  \Gamma^{+}_{{\rm{F}\rm{F}}j''}(k)\sim k^{-6}
  \left|V_{j''{\rm{FF}}}\right|^{2} \sim
  \begin{dcases}k^{0}&
    \text{if}\; j''=\rm{LA},\\
    k^{2}& \text{if}\; j''=\rm{TW},
  \end{dcases}
\end{align}
where it is possible to reapply the long-wavelength approximations of
$V_{\rm{LFF}}$ and $V_{\rm{TFF}}$ since
$V_{jj'j''}(k,\pm k',-k\mp k')$ is symmetric under exchange of phonon
indices
$(j,k)\leftrightarrow (j,\pm k') \leftrightarrow (j'',-k\mp k')$
\cite{Born1954}.

For strained tubes, $\epsilon >0$, we find
\begin{align}
  \label{eq:37}
  k^*= \frac{\sqrt{\mathcal{T}_1\epsilon}-v_{j''}}{\sqrt{\mathcal{T}_1\epsilon}+v_{j''}}\,k\quad\text{and}\quad\left|\frac{\partial
  \Omega^{+}(k,k^*)}{\partial k'}\right|=\sqrt{\mathcal{T}_1\epsilon}+v_{j''}
\end{align}
and transition rates according to Eq.~\eqref{eq:7} scale as
\begin{align}
  \label{eq:23}
  \Gamma^{+}_{{\rm{F}\rm{F}}j''}(k)\sim \epsilon^{-2} k^{-4}
  \left|V_{j''{\rm{FF}}}\right|^{2} \sim
  \begin{dcases}\epsilon^{-1 }k^{2}&
    \text{if}\; j''=\rm{LA},\\
    \epsilon^{-1} k^{4}& \text{if}\; j''=\rm{TW}.
  \end{dcases}
\end{align}

\subsection{One longitudinal and two twisting modes}
\label{sec:onel-two-twist}

In a similar manner, the strain independent scaling relations
\begin{align}
  \label{eq:32}
  \Gamma^{-}_{\rm{LTT}}(k)\sim k^2\quad\text{and}\quad
  \Gamma^{+}_{\rm{TTL}}(k)\sim k^2
\end{align}
can be inferred from Eq.~\eqref{eq:7} by noting that
$V_{\rm{LTT}}\sim kk'k''$.

\subsection{Absorption by optical phonons}
\label{sec:absorpt-optic-phon}

If a low-energy acoustic mode $(j,k)$ is absorbed by two high-energy
optical phonons, energy conservation implies
\begin{align}
  \label{eq:45}
  \Omega^{+}(k,k^*)=\omega_{j}(k)+\omega_{\rm{OPT}_a}(k^*)-\omega_{\rm{OPT}_b}(k^*+k)=0.
\end{align}
Such transitions manifest at some finite wave number $k^*\sim const.$
either in the form of intra-, $\rm{a}=\rm{b}$, or interbranch
transitions, $\rm{a}\neq \rm{b}$, where two optical branches
cross. For $ka\ll1$, we can expand $\omega_{\rm{OPT}_b}$ around $k^*$
to show that
\begin{align}
  \label{eq:25}
  \left|\frac{\partial \Omega^{+}(k,k^*)}{\partial
  k'}\right|=\left|\frac{\partial \omega_{\rm{OPT}_a}(k^*)}{\partial
  k'}-\frac{\partial \omega_{\rm{OPT}_b}(k^*)}{\partial
  k'}-\frac{\partial^2 \omega_{\rm{OPT}_b}(k^*)}{\partial k'^2}\;k-\dots\right|\sim  \begin{dcases}k&
\text{if}\; \rm{a}=\rm{b},\\
const.& \text{if}\; \rm{a}\neq\rm{b},
  \end{dcases}
\end{align}
and it follows from Eq.~\eqref{eq:7} that transition rates scale as
\begin{align}
  \label{eq:23}
  \Gamma^{+}_{j{\rm{O}_a\rm{O}_b}}(k)\sim
  \begin{dcases}
    k^{-1} \left|V_{j\rm{O}\rm{O}}\right|^2&\text{if}\; \rm{a}=\rm{b},\\
    \phantom{k^{-1}} \left|V_{j\rm{O}\rm{O}}\right|^2 &\text{if}\;
    \rm{a}\neq\rm{b}.
  \end{dcases}
\end{align}
Three-phonon coupling coefficients for one long-wavelength acoustic
$(j,k)$ and two optical phonons $V_{j\rm{O}\rm{O}}$ are of at
least linear order in $k$~\cite{Pitaevskii1981}. Our data in
Fig.~\ref{fig:3} for optical interbranch transitions of FA modes
suggests $V_{\rm{FOO}}\sim k^2$, whereas the scaling behavior of
TW modes as shown in Fig.~\ref{fig:4} suggests
$V_{\rm{TOO}}\sim k$ for both intra- and interbranch
transitions.

\section{Asymptotic strain and tube-length dependence of
  $\kappa^{\rm{RTA}}$}
\label{sec:asympt-stra-tube}

Under the RTA, the conductivity contribution $\kappa_j$ of
long-wavelength acoustic modes in tubes of length $L$ follows from
Eqs.~\eqref{eq:5} and~\eqref{eq:6} as
\begin{align}
  \label{eq:41}
  \kappa_j \sim
  L\int_{0}^{k_{\rm{cut}}}{\rm{d}}k\,\frac{\xi_1^2k^{2t}}{\xi_{2}L k^s+\xi_1k^t},
\end{align}
where it is assumed that $n=k_{\rm{B}}T/\hbar \omega$, $v=\xi_1k^t$,
and $\Gamma=\xi_2k^s$ hold true below some finite wave number
$k_{\rm{cut}}$. For linear and quadratic dispersion branches, we have
$t=0$ and $t=1$, respectively. The expression~\eqref{eq:41} is
divergent for $L\rightarrow \infty$ if $s\geq 2t+1$. This is the case
for stretched tubes, $\epsilon>0$, where we have $s\in\{1,2\}$ and
$t=0$.  The integral is then straightforwardly evaluated with strain
dependent coefficients $\xi_1$ and $\xi_2$, yielding the long-tube
scaling relations as described in Fig.~\ref{fig:6}.

\twocolumngrid

\bibliography{ppr2}

\end{document}